\documentclass[graybox]{svmult}

\usepackage{mathptmx}       % selects Times Roman as basic font
\usepackage{helvet}         % selects Helvetica as sans-serif font
\usepackage{courier}        % selects Courier as typewriter font
\usepackage{type1cm}        % activate if the above 3 fonts are
\usepackage{makeidx}         % allows index generation
\usepackage{graphicx}        % standard LaTeX graphics tool
\usepackage{multicol}        % used for the two-column index
\usepackage[bottom]{footmisc}% places footnotes at page bottom
\usepackage{latexsym,amsmath,amsfonts,amssymb}

\usepackage{epsfig}

\makeindex             % used for the subject index
\begin{document}

\title*{Midrapidity Hyperon Production in pp\\ and pA Collisions}
\author{G.H. Arakelyan, C. Merino, and Yu.M. Shabelski}
\institute{G.H. Arakelyan \at A.Alikhanyan National Scientific Laboratory
(Yerevan Physics Institute),
Yerevan, 0036, Armenia. \email{argev@mail.yerphi.am} \and C. Merino \at
Departamento de F\'\i sica de Part\'\i culas, Facultade de F\'\i sica
and Instituto Galego de F\'\i sica de Altas Enerx\'\i as (IGFAE),
Universidade de Santiago de Compostela,
Santiago de Compostela 15782,
Galiza, Spain. \email{carlos.merino@usc.es} \and
Yu.M. Shabelski \at Petersburg Nuclear Physics Institute,
NCR Kurchatov Institute,
Gatchina, St.Petersburg 188350, Russia. \email{shabelsk@thd.pnpi.spb.ru}}

\maketitle

\begin{center}
\vspace{-2.5cm}
{\it Presented by C.~Merino at Initial Stages 2014 (IS2014)}
\vskip 0.15 truecm

{\it The 2nd International Conference on the}

{\it Initial Stages in High-Energy Nuclear Collisions}
\vskip 0.15 truecm

{\it Napa Valley, CA, USA, 3-7 December, 2014}
\vskip 1.25 truecm

\end{center}

\abstract{In this paper we present the description of the production
of strange and mulistrange baryons in a wide energy region, from CERN SpS
up to LHC, in the framework of the Quark-Gluon String model.}
\vskip 1.cm

The Quark-Gluon String model (QGSM)~\cite{KTM,K20,ACKS}
is built on the Dual Topological Unitarization, nonperturbative
notions of QCD, and Regge Theory based phenomenology.
The QGSM has been successfully
used for the description of multiparticle production processes in hadron-hadron~\cite{AMPS,MPS} 
and hadron-nucleus~\cite{KTMS} collisions. In the case of interaction with a nuclear target,
the Multiple Scattering Theory (Gribov-Glauber Theory) is implemented.

Though it is not the direct aim of the present paper,
one has to note that such an approach based on the analysis of Feynman and reggeon
diagrams has been already used~\cite{BKK1,BKK2} to study the anysotropic flows, and,
in particular, the elliptic flow $v_2$ in collisions of hadrons
and nuclei at high energies, questions so extensively discussed during this conference.

The significant differences observed in the yields of baryons and antibaryons in the central
(midrapidity) region are connected with~\cite{ACKS,MRS} 
the special structure of baryons, consisting of three valence quarks together with 
a special configuration of the gluon field, called String Junction~\cite{RV}.

At very high energies, the contribution of the enhancement Reggeon diagrams
becomes important, leading to the suppression of the inclusive density of secondaries~\cite{CKTr}
in the central (midrapidity) region.

The relative probabilities for the production of various baryons depend on the
universal suppression factor S/L, for which we take the value, S/L = 0.32~\cite{AKMS},
and they can be found on the basis of simple combinatorial analysis of quarks~\cite{AnSh,CS}.

In QGSM the inclusive spectrum of a secondary hadron $h$ is determined by the
convolution of the diquark, valence quark, and sea quark distributions,
$u(x,n)$, in the incident particles, with the fragmentation functions, $G^h(z)$,
of quarks and diquarks into the secondary hadron $h$~\cite{KTM,K20}. Both the distributions
and the fragmentation functions are constructed by using the Reggeon counting rules~\cite{Kai}.

For the case of interaction with a nucleus, it is technically more simple~\cite{MPS1,pPbold} 
to consider the maximal number of Pomerons, $n_{max}$, emitted by one nucleon in the central 
region that can be cut. In this frame, we obtain a reasonable agreement with the experimental data 
on the inclusive spectra of secondaries produced in p+Pb collisions at LHC energy~\cite{pPbold}
with the value $n_{max} = 23$. It has been shown in~\cite{JDDCP} that the number of strings that
can be used for the secondary production should increase with the initial energy. 

The QGSM fragmentation formalism allows one to 
calculate the integrated over $p_T$ spectra of different secondaries as the
functions of rapidity and $x_F$, the accuracy of these calculations being of about $10\%$.

In Table~1, and in figs.~1 and 2, we compare the existing experimental data
on the energy dependence until LHC energies integrated over the whole range of $p_T$
for $\Lambda$, $\overline{\Lambda}$, $\Xi^-$, and  $\overline{\Xi}^+$ hyperons production 
density, dn/dy $(\mid y\mid \leq 0.5)$, in pp
(upper panel in Fig.~1)~[19-27]
and pA [28-31] (lower panel in Fig.~1, and Fig.~2) 
collisions to the corresponding results obtained by the QGSM. 

In Fig. 1 we also present the QGSM prediction for the energy dependence of the
$\Omega^-$ and $\overline{\Omega}^+$
hyperon production until LHC energies.

As one can see in Fig.~1, the absolute values of the densities $dn/dy(\mid y\mid \leq 0.5)$ for
$\Lambda$, $\overline{\Lambda}$ production are one order of magnitude higher than
for $\Xi^-$ and $\overline{\Xi}^+$ in a large energy region up to the LHC range.
The same is also true for $\Omega^-$ and $\overline{\Omega}^+$ densities when compared to
those of $\Xi^-$ and $\overline{\Xi}^+$.

\begin{center}
\vskip -5pt
\begin{tabular}{|c|c|c|c|c|c|c|}
\hline

$\sqrt{s}$ (GeV) & Reaction   & QGSM & Experiment dn/dy $(\mid y\mid \leq 0.5)$  \\
\hline

13.97 (102 GeV/c) & p + p $\to \Lambda$ & 0.025& $0.012 \pm 0.01$ \cite{chap}  \\
\hline

14.075 (147 GeV/c) & p + p $\to \Lambda$ & 0.025 & $0.0090 \pm 0.0015$ \cite{brick}  \\  

14.075 (147 GeV/c) & p + p $\to \overline{\Lambda}$ & 0.010 & $0.057 \pm 0.0044$ \cite{brick}  \\  \hline

19.42 (200 GeV/c) & p + p $\to \Lambda$ & 0.026 & $0.0106 \pm 0.006$ \cite{lopinto}  \\  

19.42 (200 GeV/c) & p + p $\to \overline{\Lambda}$ & 0.011 & $0.007 \pm 0.0015$ \cite{lopinto}  \\  \hline

19.66 (205 GeV/c) & p + p $\to \Lambda$ &0.026 & $0.015 \pm 0.0044$ \cite{jaeger}  \\  \hline

23.76 (300 GeV/c) & p + p $\to \Lambda$ & 0.026 & $0.0076 \pm 0.006$ \cite{sheng}  \\  

23.76 (300 GeV/c) & p + p $\to  \overline{\Lambda}$ & 0.013 & $0.015 \pm 0.0076$ \cite{sheng}  \\  \hline

27.6 (405 GeV/c) & p + p $\to \Lambda$ &0.027 & $0.0091 \pm 0.0045$ \cite{kichimi}  \\  

27.6 (405 GeV/c) & p + p $\to \overline{\Lambda}$ & 0.015& $0.017 \pm 0.0041$ \cite{kichimi}  \\  \hline

200. & p + p $\to \Lambda$ & 0.045 & $0.0436 \pm 0.0008 \pm 0.004$ \cite{STAR0}  \\  

200. & p + p $\to \overline{\Lambda}$ & 0.039 & $0.0398 \pm 0.0008 \pm 0.0037$ 
\cite{STAR0} \\ \hline

200. & p + p $\to \Lambda(FD)$ & 0.045 & $0.0385 \pm 0.0007 \pm 0.0035$ \cite{STAR0} \\  

200. & p + p $\to \overline{\Lambda}(FD)$ & 0.039 & $0.0351 \pm 0.0007 \pm 0.0032$ 
\cite{STAR0} \\  \hline

200. & p + p $\to \Xi^- $ & 0.005 & $0.0026 \pm 0.0002 \pm 0.0009$ \cite{STAR0} \\   

200. & p + p $\to \overline{\Xi}^+$ & 0.005 & $0.0029 \pm 0.0003 \pm 0.006$ 
\cite{STAR0} \\ \hline

200. & p + p $\to \Omega^- + \overline{\Omega}^+ $ & 0.001 & $0.00034 \pm 0.00016 \pm 0.0005$ 
\cite{STAR0} \\  \hline

900. & p + p $\to \Lambda$ & 0.065 & $0.26 \pm 0.01$ \cite{ATLAS} \\  \hline

7000. & p + p $\to \Lambda$ & 0.099 & $0.27 \pm 0.01$ \cite{ATLAS} \\ \hline

900. & p + p $\to \Lambda$ &0.065 & $0.108 \pm 0.001 \pm 0.012$ \cite{CMS} \\ \hline

7000. & p + p $\to \Lambda$ & 0.099 & $0.189 \pm 0.001 \pm 0.022$ \cite{CMS} \\  \hline

900. & p + p $\to \Xi^-$ & 0.009 & $0.011 \pm 0.001 \pm 0.001 $ \cite{CMS} \\ \hline

7000. & p + p $\to \Xi^-$ & 0.015 & $0.021 \pm 0.001 \pm 0.003 $  \cite{CMS} \\

\hline\end{tabular}
\end{center}
{\footnotesize {\bf Table 1:} Experimental data for strange baryons and antibaryons production
in pp collisions at different energies, from CERN SpS up to LHC, together
with the corresponding description by the QGSM.}
\vskip 0.5cm

\begin{figure}[htb]
\centering
\vskip -4.cm
\includegraphics[width=0.8\hsize]{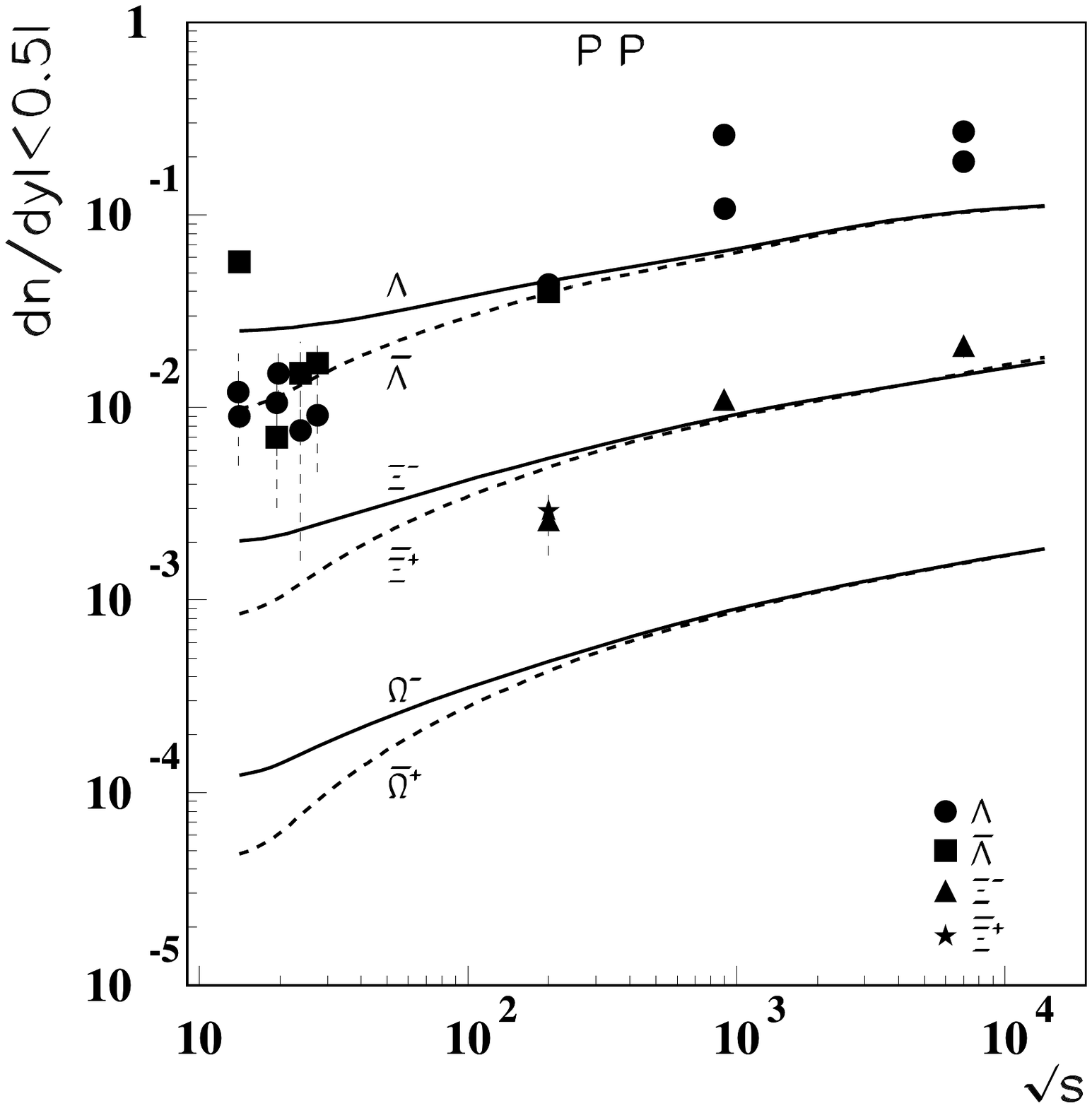}
\vskip -4.5cm

\includegraphics[width=0.8\hsize]{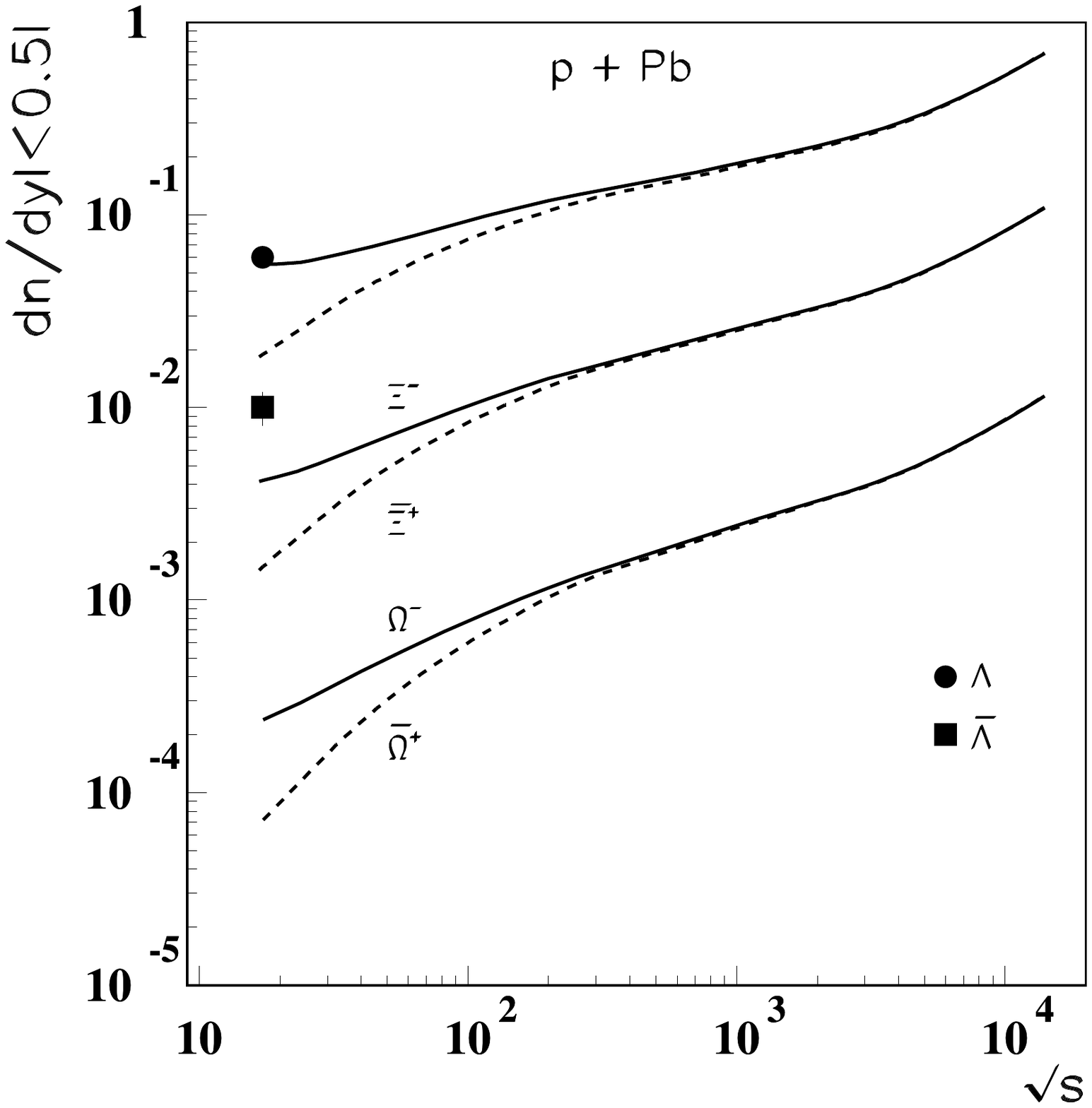}
\vskip -.4cm
\caption{\footnotesize
Experimental data, integrated over $p_T$, on the energy dependence of hyperon production
in midrapidity region for pp (upper panel) and p+Pb (lower panel)
collisions, compared to the corresponding QGSM calculations (baryons are shown by full curves
and antibaryons by dashed curves).}
\end{figure}

\begin{figure}[htb]
\centering
\vskip -4.cm
\includegraphics[width=.8\hsize]{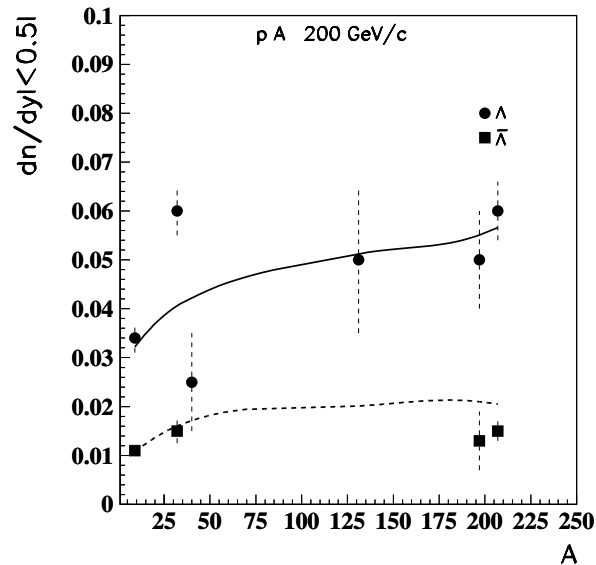}
\vskip -.2cm
\caption{\footnotesize
Experimental data, integrated over $p_T$, on the A-dependence of the midrapidity denstity
of $\Lambda$ and $\overline{\Lambda}$ hyperons produced at 200 GeV/c~\cite{NA5,NA35a,NA35b,NA57},
compared to the corresponding QGSM calculations
(baryons are shown by full curves and antibaryons by dashed curves).}
\end{figure}

\begin{center}
\vskip -5pt
\begin{tabular}{|c|c|c|c|c|c|c|} \hline

$\sqrt{s}$ (GeV) & Reaction   & QGSM & Experiment  dn/dy($\mid y\mid \leq 0.5$)  \\
\hline

19.42 (200 GeV/c) & p + Ar $\to \Lambda$ & 0.042 & $0.025 \pm 0.01 $ \cite{NA5} \\
\hline

19.42 (200 GeV/c)& P + Xe $\to \Lambda$ & 0.052 & $0.05 \pm 0.015 $ \cite{NA5} \\
\hline

19.42 (200 GeV/c) & p + Au $\to \Lambda$ & 0.055 & $0.05 \pm 0.01 $ \cite{NA35a} \\

19.42 (200 GeV/c) & p + Au $\to \overline{\Lambda}$ & 0.021 & $0.013 \pm 0.06 $ \cite{NA35a} \\
\hline

19.42 (200 GeV/c) & p + S $\to \Lambda$ & 0.041 & $0.06 \pm 0.005 $ \cite{NA35b} \\

19.42 (200 GeV/c) & p + S $\to \overline{\Lambda}$ & 0.016 & $0.015 \pm 0.0025 $ \cite{NA35b} \\
\hline

17.2 & p + Be $\to \Lambda$ & 0.033 & $0.034 \pm 0.0005 \pm 0.003$ \cite{NA57} \\

(m.b.) & p + Be $\to \overline{\Lambda}$ & 0.011 & $0.011 \pm 0.0002 \pm 0.001$ \cite{NA57} \\ 
\hline

17.2 & p + Pb $\to \Lambda$ & 0.055 & $0.060 \pm 0.002 \pm 0.006$ \cite{NA57} \\

(m.b.) & p + Pb $\to \overline{\Lambda}$ & 0.019 & $0.015 \pm 0.001 \pm 0.002$ \cite{NA57} \\ \hline
\end{tabular}
\end{center}
{\footnotesize {\bf Table 2:} Experimental data for strange baryons and antibaryons production
in proton-nucleus collisions at different energies, together with the
corresponding description by the QGSM.}
\vskip 0.4cm

As one can see, both in figs.~1 and 2 and in Table~1, the experimental data on
$dn/dy$$(\mid y\mid \leq 0.5)$ obtained by different collaborations are not
thoroughly compatible among them, what it is probably due to different experimental
event selections, especially when comparing measurements by CMS and ATLAS collaborations.
Nonetheless, the agreement of the QGSM results with the different sets of experimental data
is good enough when compared to those experimental discrepancies. 

In Fig. 2 the A-dependences of $\Lambda$ and $\overline{\Lambda}$ hyperons
produced on nuclear targets~\cite{NA5,NA35a,NA35b,NA57} are shown. Also here
the QGSM curves are in a reasonable agreement with the experimenatl data.
\vskip 0.75cm

{\small\bf Acknowledgements}

C.M. wants to congratulate the organizers for getting the scientifically exciting
environment that it has made this conference so fruitful.

We thank C. Pajares for useful discussions.
This paper was supported by Ministerio de Econom\'ia y
Competitividad of Spain (FPA2011$-$22776), the Spanish
Consolider-Ingenio 2010 Programme CPAN (CSD2007-00042),
by Xunta de Galicia, Spain (2011/PC043), by the State
Committee of Science of the Republic of Armenia
(Grant-13-1C023), and by Russian RSCF grant No.14-22-00281.

\end{document}